\documentclass[journal,twoside,web]{ieeecolor}
\usepackage{tmi}
\usepackage{cite}
\usepackage{amsmath,amssymb,amsfonts}
\usepackage{algorithmic}
\usepackage{graphicx}
\usepackage{textcomp}
\usepackage{algorithm2e}
\usepackage{booktabs,ragged2e}
\usepackage[flushleft]{threeparttable}

\makeatletter
\g@addto@macro\@afterheading{\normalcolor}
\makeatother

\def\BibTeX{{\rm B\kern-.05em{\sc i\kern-.025em b}\kern-.08em
    T\kern-.1667em\lower.7ex\hbox{E}\kern-.125emX}}
\begin{document}

\bstctlcite{BSTcontrol}

\title{vPET-ABC: Fast Voxelwise Approximate Bayesian Inference for Kinetic Modeling in PET}
\author{Qinlin Gu$^\ast$ \IEEEmembership{Student Member, IEEE}, Gaelle M. Emvalomenos$^\ast$, Evan D. Morris, Clara Grazian, Steven R. Meikle \IEEEmembership{Senior Member, IEEE}
\thanks{$^\ast$\emph{Q. Gu and G. M. Emvalomenos contributed equally to this work.}}
\thanks{This work is supported by the Australian Research Council (Discovery Projects grant DP230102070).}
\thanks{Qinlin Gu (qinlin.gu@sydney.edu.au) is with the School of Mathematics and Statistics and the Brain and Mind Centre, the University of Sydney, Australia.}
\thanks{Gaelle M. Emvalomenos is with Radiology and Biomedical Imaging, Yale University, USA.}
\thanks{Evan D. Morris is with Radiology and Biomedical Imaging, Biomedical Engineering, and Psychiatry, Yale University, USA.}
\thanks{Clara Grazian is with the School of Mathemtatics and Statistics and the ARC Training Centre in Data Analytics for Resources and Environments (DARE), the University of Sydney, Australia.}
\thanks{Steven R. Meikle is with the School of Health Sciences, Sydney Imaging and the Brain and Mind Centre, the University of Sydney, Australia.}
}
\maketitle


\begin{abstract}

Dynamic PET kinetic modeling increasingly demands voxelwise uncertainty quantification and robust model selection. Yet total-body PET (TB-PET) data volumes make conventional Bayesian approaches, such as per-voxel MCMC, computationally impractical, while deep models typically require retraining and careful revalidation when tracers, protocols, or kinetic models change, without necessarily improving inference speed.

Vectorized voxelwise approximate Bayesian computation (vPET-ABC) is introduced as a likelihood-free, model-agnostic posterior inference framework for dynamic PET kinetic modeling at total-body scale. The method replaces explicit likelihood evaluation with forward simulations and a discrepancy test, then exploits full vectorization to transform voxelwise inference into an embarrassingly parallel workload suited to modern GPUs.

In simulation, vPET-ABC produced posterior summaries with small divergence from sequential Monte Carlo baselines, and posterior mean estimates significantly more accurate than non-negative least squares (NNLS). For model selection between the linear parametric neurotransmitter model (lp-ntPET) and the multilinear reference tissue model, vPET-ABC maintained high sensitivity under high noise with moderate loss of specificity, whereas NNLS\,+\,Bayesian information criteria exhibited the opposite trade-off with near-zero sensitivity. In a proof-of-concept human cigarette smoking dataset, vPET-ABC yielded denser probabilistic activation maps than lp-ntPET with effective number of parameters. On a proof-of-concept $50$\,min total-body [$^{18}$F]FDG study, vPET-ABC generated whole volume $K_i$ parametric images within practical runtimes on a single data\textendash center GPU. The results aligned with PET SUV, handled voxels violating irreversibility, and preserved local spatial correlation better than NNLS.

Overall, vPET-ABC delivers fast, training-free, uncertainty-aware inference that scales to TB-PET and remains portable across tracers and kinetic models.
\end{abstract}


\section{Introduction}
\label{sec:introduction}

\IEEEPARstart{W}{hen} applied to dynamic PET, kinetic modeling enables noninvasive quantification of physiological and molecular processes from time–activity curves (TACs) \cite{wang_pet_2020}. It describes radiotracer transport, tissue uptake and target engagement using compartmental models in the form of ordinary differential equations. This general methodology can be applied to a wide range of clinical and preclinical applications ranging from neurotransmitter studies in the neuroscience of addiction to tumor characterization in oncology \cite{wang_pet_2020}. Reliable estimation and inference are therefore central to PET quantification.

Total-body PET (TB-PET) is a recent advance whereby the axial field-of-view of the scanner is extended to 1 to 2 meters, resulting in ultra-high sensitivity \cite{cherry_total-body_2018}, which improves signal-to-noise relative to injected dose and enables simultaneous whole-body kinetic analysis. However, it also introduces a model-selection challenge due to the wide variation in tracer kinetic behavior throughout the body and imposes a prohibitive computational burden for voxelwise parameter and uncertainty estimation at scale \cite{wang_total-body_2022}. Against this backdrop, we summarize existing approaches and the practical trade-offs that shape current PET kinetic analysis methods.

In total-body PET, fast semi-quantitative whole-body parametric imaging is often performed with Patlak graphical analysis \cite{patlak_graphical_1983} to obtain voxelwise $K_i$ (net influx). However, standard Patlak relies on the assumption of irreversibile trapping of the tracer which may be valid in some organs but not others, motivating generalized Patlak formulations \cite{karakatsanis_generalized_2015}. Patlak-style methods also focus primarily on $K_i$, while full compartmental modeling can provide additional macro- and micro-kinetic parameters that enable multiparametric characterization in total-body applications \cite{wang_total-body_2022}.

Deterministic estimators, such as fast non-negative least squares (FNNLS), are widely used in PET compartmental modeling, especially in basis-function formulations of kinetic models \cite{gunn_positron_2002}. Although simple and fast, least-squares basis-function approaches can be sensitive to observational noise and their accuracy depends on basis-function choice, which can reduce sensitivity and reliability in practice \cite{irace_bayesian_2020}. These limitations become especially consequential when the goal is not only parameter estimation but also deciding between competing kinetic models, since the decision is made under the same noise, identifiability, and constraint structure.

Bayesian approaches address these issues by providing a Monte Carlo approximation to the full posterior probability distribution, enabling a more complete picture of parameter uncertainty \cite{de_gier_approximate_2019}. The feasibility of posterior estimation in dynamic PET has been demonstrated at the region of interest (ROI) level, including Bayesian neurotransmitter PET (b-ntPET) \cite{irace_bayesian_2020} implemented with Markov chain Monte Carlo (MCMC) based on the linear parametric neurotransmitter PET model (lp-ntPET) \cite{normandin_linear_2012}, and likelihood-free approximate Bayesian computation for PET kinetic modeling (PET-ABC) \cite{fan_pet-abc_2021}. However, conventional Monte Carlo methods \cite{metropolis_equation_1953} are difficult to scale voxelwise due to sequential sampling and convergence variability, while recent deep generative approaches tend to be slow at inference time (approx. 1 min/TAC) and still typically require retraining and revalidation when the tracer, protocol, or model changes \cite{djebra_bayesian_2025}.

We, therefore, propose vectorized voxelwise approximate Bayesian computation for PET kinetic modeling (vPET-ABC), a fast, likelihood-free, model-agnostic posterior sampler. This method builds on our prior work on PET-ABC\cite{fan_pet-abc_2021, grazian_vpet-abc_2021} by leveraging parallel simulation to deliver voxelwise posterior inference with calibrated uncertainty at whole-body scale.

In this study, we (i) formulate vPET-ABC for dynamic PET kinetic modeling and validate conditions under which vectorized, likelihood-free inference recovers accurate posteriors, (ii) provide a GPU-accelerated end-to-end pipeline from time activity curves (TACs) to posterior summaries that scales to whole-body volumes, (iii) evaluate the performance of vPET-ABC on two kinetic-model simulation benchmarks, assessing point-estimate accuracy, posterior fidelity, and model selection sensitivity, and (iv) demonstrate deployment in clinical studies without the requirement of training across tracers, models, and acquisition protocols. By combining likelihood-free Bayesian inference with vectorized computation, we show that vPET-ABC offers a practical, uncertainty-aware framework for large-scale kinetic parameter estimation aligned with contemporary TB-PET data volumes and hardware.

\section{Methods}
\label{sec:methods}

\subsection{Kinetic Models}

Many kinetic models have been proposed over decades of research for a wide variety of tracers and molecular targets \cite{morris_kinetic_nodate}. Here, we evaluate the performance and versatility of vPET-ABC in two distinct contexts: (i) assessing the reversibility or irreversibility of [$^{18}$F]fluorodeyglucose (FDG) using the two tissue compartment model (2TCM), and (ii) analyzing stimulus-evoked neurotransmitter release in response to an external stimulus during scanning using the linearized parametric neurotransmitter PET model (lp-ntPET)\cite{normandin_linear_2012} as in \cite{morris_modeling_2024}.

\subsubsection{Two Tissue Compartment Model}

The two-tissue compartment model \cite{phelps_tomographic_1979} represents tracer kinetics with a plasma input $C_p$ and two tissue compartments in series: free tracer $C_f$ and metabolized tracer $C_m$. Compartment exchange is governed by non-negative rate constants $K_1$ (plasma$\to C_f$), $k_2$ ($C_f\to$ plasma), $k_3$ ($C_f\to C_m$), and $k_4$ ($C_m\to C_f$), with $k_4=0$ indicating irreversible trapping. The dynamics are described by
\begin{align}
\label{eq:2TCM}
\frac{\mathrm{d}C_f}{\mathrm{d}t} &= K_1 C_p - (k_2 + k_3)C_f + k_4 C_m,\nonumber\\
\frac{\mathrm{d}C_m}{\mathrm{d}t} &= k_3 C_f - k_4 C_m,
\end{align}
and the measured PET activity is
\begin{equation}
\label{eq:2TCM_op}
C_t(t) = (1-V_b)\big(C_f(t)+C_m(t)\big)+V_b C_{wb}(t),
\end{equation}
where $V_b\in(0,1)$ is the blood volume fraction and $C_{wb}$ is whole-blood activity. In the irreversible case ($k_4=0$), the model conforms to the Patlak linearization \cite{patlak_graphical_1983}, although reversibility can be tissue-dependent \cite{karakatsanis_generalized_2015}. For [\textsuperscript{18}F]FDG, we assume $C_p \approx C_{wb}$.

\subsubsection{Linear Parametric Neurotransmitter PET model}

The lp-ntPET model \cite{normandin_linear_2012} is a reference-tissue based non-steady state model for detecting transient task- or drug-evoked neurotransmitter changes without arterial sampling. It extends the multilinear reference tissue model (MRTM) \cite{ichise_linearized_2003} by adding a time-varying activation term that captures transient ligand displacement, typically parameterized using a basis set spanning a range of plausible onset, peak, and decay profiles. The linearized form is
\begin{equation}
\label{eq:lp-ntPET}
C_t(t)=R_1C_r(t)+k_2\int_0^tC_r(u)\mathrm{d}u-k_{2a}\int_0^tC_t(u)\mathrm{d}u-\gamma B(t),
\end{equation}
with
\begin{equation}
\label{eq:Bt}
B(t)=\int_0^tC_t(u)g(u)\mathrm{d}u,
\end{equation}
where $C_r(t)$ is the reference-region TAC, $R_1=\frac{K_1}{K_1^{\prime}}$ is the relative tracer delivery to target vs.\ reference, $k_2$ is the target-tissue efflux rate constant (tissue$\rightarrow$plasma), $k_{2a}=\frac{k_2}{1+BP_{ND}}$ is the apparent efflux rate constant that incorporates specific binding, and $g(t)$ is a gamma-variate activation profile \cite{normandin_linear_2012}. Setting $\gamma=0$ reduces the model to MRTM, so lp-ntPET can be viewed as a generalization of MRTM.

\subsection{Bayesian Statistics}

In kinetic modeling, Bayes’ rule updates beliefs about parameters (e.g., $\boldsymbol{\theta}=(R_1,k_2,k_{2a},\ldots)^{\mathsf T}$) given observed TACs $\mathbf{C}$ (e.g., $\{C_t(t_i)\}_{i=1}^n$ and optionally $C_r(t)$):
\begin{equation}
\label{eq_Bayes_theorem_para}
\pi(\boldsymbol{\theta}\mid \mathbf{C})
=\frac{p(\mathbf{C}\mid \boldsymbol{\theta})\,\pi(\boldsymbol{\theta})}{p(\mathbf{C})}
\propto p(\mathbf{C}\mid \boldsymbol{\theta})\,\pi(\boldsymbol{\theta}).
\end{equation}

Model selection introduces a discrete variable $m\in\{1,\dots,M\}$ indexing candidate kinetic models, giving
\begin{align}
\label{eq_Bayes_theorem_para_with_model}
\pi(m,\boldsymbol{\theta}\mid \mathbf{C}) &\propto p(\mathbf{C}\mid \boldsymbol{\theta},m)\,\pi(\boldsymbol{\theta}\mid m)\,\pi(m),\nonumber\\
\pi(m\mid \mathbf{C})&=\int \pi(m,\boldsymbol{\theta}\mid \mathbf{C})\,\mathrm d\boldsymbol{\theta}.
\end{align}

When the likelihood function $p(\mathbf{C}\mid\boldsymbol{\theta})$ is tractable and a known noise model exists, Monte Carlo-based simulation methods are adopted to approximate the posterior distribution in Eq.~\eqref{eq_Bayes_theorem_para_with_model}.

As an alternative, approximate Bayesian computation (ABC) \cite{rubin_bayesianly_1984} provides likelihood-free Bayesian inference by replacing explicit likelihood evaluation with forward simulations and a discrepancy criterion, enabling posterior approximation and model selection under suitable conditions.

\subsection{Approximate Bayesian Computation and vPET-ABC}
\subsubsection{Approximate Bayesian Computation}

One of the simplest ABC algorithms is the rejection sampling \cite{Jonathan_ABC_rej} method, which approximates the likelihood function by:
\begin{small}
\begin{equation} p(\boldsymbol{\theta}\mid S(\mathbf{x}_{obs}))=\lim_{h\to0}\int_{\mathbf{X}}p(\boldsymbol{\theta},\mathbf{x}\mid\rho(S(\mathbf{x}), S(\mathbf{x}_{obs})\leq h)\mathrm{d}\mathbf{x},
    \label{eq_rejection_ABC}
\end{equation}
\end{small}
where $\mathbf{x}\in\mathbf{X}$ are simulated data, $\mathbf{x}_{obs}\in\mathbf{X}$ are observed data, $\rho(\cdot)$ is a distance function, $S(\cdot)$ is a (possibly sufficient) summary statistic to reduce the dimensionality and computation required, and $h$ is the acceptance threshold.

For each voxel, rejection ABC repeatedly samples a candidate model and its parameters from the prior, simulates a TAC, and accepts the draw if the discrepancy between simulated and observed summary statistics is below a tolerance $h$. This is repeated until $n$ draws are accepted.


\subsubsection{vPET-ABC}


\textit{vPET-ABC:} The conventional ABC algorithm can also be implemented under a fixed simulation budget $N$ (i.e., a fixed computational budget or total number of simulations). Instead of running the algorithm until $n$ draws with distances below the threshold $h$ have been accepted, one performs $N$ simulations and retains the closest $p=n/N$ proportion of simulated distances, following Biau \textit{et al.}~\cite{biau_new_2015}.



For ease of parallelization and vectorization to enable large-scale voxelwise PET kinetic modeling, we further adapted the conventional ABC algorithm in a PET kinetic modeling context to be fully compatible with matrix calculations for ease of GPU implementation, as shown in Alg. \ref{alg_vvABC}. We will show in the next section that through vectorization and the utilization of modern GPUs, we mitigate ABC’s low acceptance/space-exploration efficiency in comparison with MCMC or sequential Monte Carlo (SMC), achieving speed comparable to FNNLS. We will also show that such adaptation does not retard convergence rate.

\RestyleAlgo{ruled}
\SetKwComment{Comment}{/* }{ */}
\SetKw{KwBy}{by}
\begin{algorithm}
\caption{Vectorized Voxelwise Rejection ABC algorithm (vPET-ABC), $M$ is the number of models, $P$ is the number of parameters, $N$ is the simulation size, $J$ is the number of voxels, $p=\frac{\hat{n}}{N}$ is the proportion of accepted samples where $\hat{n}$ is the desired accepted number of samples, $L$ is the number of frames for each TAC.}
Sample model indicators $m\in\mathbf{M}=\{0,1,\dots,M-1\}$ with equal probability of $\frac{1}{M}$\;
Sample model parameters, stored together in a matrix $\boldsymbol{\Theta}$ of shape $(N, P+1)$ together with the model indicators\;
Compute model curves using $\boldsymbol{\Theta}$, stored in a matrix $\mathbf{X}$ of shape $(N, L)$\;
Compute distances on summary statistics $\rho(S(\mathbf{X}),S(\mathbf{x}_{obs}))$, stored in a matrix $\mathbf{D}$ of shape $(J,N)$\;
Compute the smallest $p$ proportion along the $N$ dimension in $\mathbf{D}$, as the accepted posteriors of all voxels with a shape of $(J,\hat{n})$\;
\label{alg_vvABC}
\end{algorithm}

Fixing $p$ in Alg. \ref{alg_vvABC} is equivalent to fixing $n$, the accepted number of samples, if $p\in[\frac{n}{\hat{N}},\frac{n+1}{\hat{N}})$ such that $n=\lfloor\hat{N}p\rfloor$ still holds.

\subsection{Optimizations and Usage Pipeline}
\label{pipeline}
This section outlines key ABC components and how to tune them, and presents a practical user-facing pipeline.

\subsubsection{Summary Statistic $S(\cdot)$}

In Fan \textit{et al.}~\cite{fan_abc_2016}, two alternative summary statistics were evaluated: the raw (noisy) observed TAC and a smoothing-spline estimate of the TAC. They observed that both performed well at low noise capturing the true curve within the $95\%$ credible interval, while the spline summary was more robust under higher noise. To mitigate the computational burden of smooth spline estimations in large datasets, we recommend using the observed TAC as the summary statistic.


\subsubsection{Threshold $h$}
The proportional threshold $p$ (acceptance rate) controls the bias–variance trade-off of ABC estimators and also affects model-selection sensitivity/specificity. 
Tuning hyperparameters $h,p,n$ is challenging, but here we propose a method inspired by Barber~\textit{et al.}\cite{barber_rate_2015}.



Tuning $h,p,n$ for parameter accuracy, we recommend a short pilot on simulation data matched in parameter ranges and noise, where the true parameter values are known: (i) for a small grid of $n$ values, compute the posterior mean of a target parameter across repetitions and estimate its MSE; (ii) for sufficiently small $h$, the MSE versus $h$ or $n$ curve is U-shaped; (iii) fit a smooth curve and identify the value corresponding to the minimum of the fitted curve.

Tuning $h$, $p$, and $n$ for model selection requires a different calibration. Under a fixed simulation budget, ABC may intrinsically penalize more complex models because their higher-dimensional parameter spaces are explored less thoroughly. By contrast, NNLS-based model selection typically relies on explicit criteria such as information criteria or $F$-tests, while ABC offers a coherent simulation-based approach to model comparison when likelihood-based criteria are unavailable or unreliable. Although recent theory for ABC model selection is available \cite{grazian_approximate_2025}, it relies on assumptions that are not satisfied in our kinetic modeling setting. We therefore recommend the following empirical calibration pipeline: (i) simulate datasets from each candidate model at representative parameter/noise settings; (ii) vary $n$ (or $p$) over a prespecified grid and compute accuracy and ROC AUC for each model; (iii) choose a working point in the elbow-shaped curve that achieves the desired balance between sensitivity and specificity across models.

\subsubsection{Point Estimate and Other Inference Questions}

In medical imaging, a common output is a parametric map that enables an informative vizualization of the kinetic modeling result. In order to obtain these images, one value per voxel is required, which is generally a frequentist point estimate. In a Bayesian setting, in order to create a parametric map, a point estimate can be the posterior mean
, the geometric median or the Minimum Mean Square Error (MMSE) applied to the Highest Posterior Density (HPD) of a region \cite{irace_bayesian_2020}. Given the theoretical guarantees and computational simplicity, we recommend the posterior mean; HPD can be tricky and computationally demanding for ABC since it requires density estimation (e.g., KDE) on accepted samples. 

\begin{figure}[!t]\centerline{\includegraphics[width=\columnwidth]{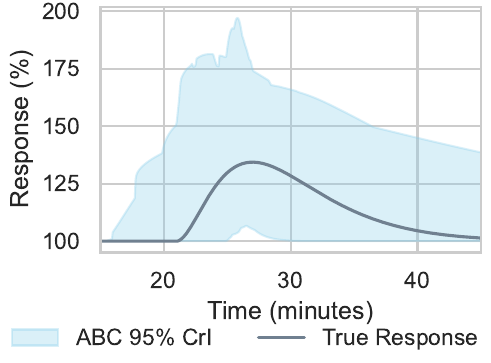}}
    \caption{$95\%$ credible interval of the response function $k_{2a}(t)/k_{2a}=1+\gamma/k_{2a}\cdot g(t)$ in the lp-ntPET model, expressed as percentage above baseline.}
    \label{fig:response}
\end{figure}

The strength of Bayesian analysis, however, lies in quantifying uncertainty via the posterior distribution. In a lp-ntPET model study, for instance, when it is difficult to decide the presence or absence of significant neurotransmitter activation, Bayesian credible interval (CrI) of the response function $\frac{k_{2a}(t)}{k_{2a}}=1+\frac{\gamma}{k_{2a}}g(t)$ provides additional insight to inform the decision. In the example of Fig.~\ref{fig:response}, we can infer activation since the lower bound of the ABC $95$\% CrI excludes a flat (null) response.

\section{Experiments}

The proposed method and pipelines were applied and evaluated on several simulation and real datasets. The two tissue compartment model simulation was used to test kinetic parameter point estimate accuracy, while the lp-ntPET/MRTM simulation was chosen to validate kinetic parameter posterior distribution accuracy, speed performance, and most importantly model selection sensitivity. Two real datasets were then selected to showcase how the method should be applied in practical cases and how well they perform compared with established non-Bayesian approaches.

\subsection{Simulation datasets}

\subsubsection{Two Tissue Compartment Model}

Ten thousand 1D 2TCM TACs were simulated from Eq.~\eqref{eq:2TCM} and Eq.~\eqref{eq:2TCM_op} using kinetic parameters \((K_{1},k_{2},k_{3},k_{4},V_{b})\) sampled uniformly from ranges reported in the PET literature for [\textsuperscript{18}F]FDG in a wide variety of organs and tissues \cite{liu2021kinetic,sari2022first}. Specifically, we chose 
\begin{align}
K_{1} &\in [0.001,\,1.00]\ \mathrm{mL\,min^{-1}\,cm^{-3}}, &k_{2} \in [0.001,\,2.00]\ \mathrm{min^{-1}},\nonumber\\
k_{3} &\in [0.001,\,0.50]\ \mathrm{min^{-1}}, &k_{4} \in [0,\,0.10]\ \mathrm{min^{-1}},\nonumber\\
V_{b} &\in [0.03,\,0.20].
\end{align}

The arterial blood input function was simulated using Feng's model \cite{feng1993models}:
\begin{equation}
C_{a}(t)=\bigl(\beta_1t-\beta_2-\beta_3\bigr)e^{-\kappa_1t}+\beta_2e^{-\kappa_2t}+\beta_3e^{-\kappa_3t}, 
\end{equation}
where the six parameters \((\beta_{1},\beta_{2},\beta_{3},\kappa_{1},\kappa_{2},\kappa_{3})\) were drawn uniformly from the following ranges:
\begin{align}
\beta_1&\in[4\times10^{4},4.7\times10^{5}],&\beta_2\in[1\times10^{4},2\times10^{5}],\nonumber\\
\beta_3&\in[8\times10^{3},3\times10^{4}],&\kappa_1\in[1,1\times10^{3}],\nonumber\\
\kappa_2&\in[0.03,3.60],&\kappa_3\in[0.01,0.035],
\end{align}
covering the full variability observed in eight oncological subjects aged 43-82 years (Mean$\pm$SD: $70.53 \pm 10.36$), weighing 55-97 Kg (Mean$\pm$SD: $71.23 \pm 12.80$ Kg) acquired on a Siemens Biograph Vision Quadra (Siemens Healthineers, Knoxville, TN). All subjects were administered [\textsuperscript{18}F]FDG as an intravenous bolus (Mean$\pm$SD: $146.67 \pm 11.50$ MBq in 1.0$\pm$0.5 ml at 10 ml/min).

Gaussian noise was then added to the simulated TACs, $C_t$, according to the model \cite{fan_pet-abc_2021}:
\begin{equation}
    y(t) = C_t(t) + \ell\,\sigma_t\,\mathcal{N}(0,1),
\end{equation}
where $\sigma_t=\sqrt{\frac{C_t(t)\mathrm{exp}(-\lambda t)}{\Delta t}}\mathrm{exp}(\lambda t)$. $\Delta t$ is the duration of a given frame in the PET time series, $\lambda=\mathrm{ln}2/T_{1/2}$ is the decay constant for $\textsuperscript{18}{}{\mathrm{F}}$, with $T_{1/2}=109.8$ min and $l$ is a constant that scales the noise level. To match realistic noise levels, $l=7$ was estimated from the aforementioned eight PET scans using MCMC.

Standard least squares method (Trust Region Reflective algorithm implemented via $\texttt{scipy.optimize.least\_squares}$ using default arguments) and vPET-ABC ($N=10,000,000, n=18$) were applied to the simulation dataset, and evaluated by comparison to ground truth.

\subsubsection{Linear Parametric Neurotransmitter PET model}

Two 1D simulation datasets at different noise levels, each containing $4000$ [$^{11}$C]raclopride TACs with $61$ frames of 60 seconds duration, were generated using the neurotransmitter PET (ntPET) model \cite{morris_ntpet_2005} with the same parameters as from the literature 
\cite{angelis_direct_2019}
. Half included dopamine activation, indicated by non-zero $\gamma$, while the remainder had no activation ($\gamma$=0). Gaussian and Poisson noise models were applied separately \cite{fan_pet-abc_2021}, each to $2000$ TAC simulations:
\begin{align}\epsilon_{\text{Gaussian}}&\sim\mathcal{N}(0,l_1\sqrt{\frac{C_T(t)}{\Delta te^{\lambda t}}}),\nonumber\\
    \epsilon_{\text{Poisson}}&\sim\frac{\text{Pois}(l_2C_T(t))}{l_2},
\end{align}
where $l_1,l_2$ are the noise-level constants and $\lambda=\ln\frac{2}{T_{1/2}}$ is the decay constant for the tracer. The two noise levels chosen correspond to noise observed on a microPET Focus $220$ (Siemens Healthineers, Knoxville, TN) (high noise) and a Bruker Si$78$ (Bruker Biospin, Ettlingen, Germany) (low noise). $l_1,l_2$ were estimated from real Bruker/microPET scans using an MCMC sampler under the corresponding noise model, and then used as constants for the simulation.

NNLS with basis functions \cite{normandin_linear_2012}, vPET-ABC, SMC \cite{ChopinPapaspiliopoulos2020SMC} and Hamiltonian Monte Carlo (HMC) \cite{hoffman_no-u-turn_2011} were applied to the simulated datasets for further evaluation of computational speed, model selection sensitivity and kinetic parameter posterior accuracy. $t_D$, $t_P$, and $\alpha$ were discretized onto predefined grids to construct the basis-function library. Specifically, $t_D\in[15,25]$ with unit increments; for each $t_D$, $t_P\in[t_D+1,,t_D+45]$ with unit increments; and $\alpha\in{0.25,1,4}$, consistent with prior work \cite{liu_detecting_2021}.

\subsection{Proof-of-Concept Studies}
\subsubsection{Human Cigarette Smoking Study}
\label{real_smoking}

To apply the vPET-ABC pipeline to real dynamic PET data for capturing stimulus-evoked neurotransmitter release, we followed the pipeline in Section~\ref{pipeline}, beginning with realistic simulations to optimize $n$ (desired number of accepted simulations) for model selection. We generated realistic 90-min TACs by creating a 4D human striatum phantom, as described by Liu and Morris\cite{liu_detecting_2021}, for which [$^{11}$C]raclopride kinetics were simulated using COMKAT in Matlab (Mathworks) \cite{Muzic636} with noise corresponding to the HRRT PET scanner (Siemens Healthineers, Knoxville, TN), then denoised with a $3\times3\times3$ voxel HighlY constrained backPRojection (HYPR) filter \cite{christian_dynamic_2010}, exactly matching the processing applied to real data. As in \cite{liu_detecting_2021}, a cerebellar reference-region TAC was simulated, along with striatal TACs spanning fast and slow kinetics: 1345 voxels with dopamine activation and 505 voxels without activation (NULL). 
Dopamine activation profiles were modeled as a gamma-variate scaled by magnitude $G$, $\mathrm{DA}(t)=Gg(t)+\text{baseline}$, where $G$ is the peak DA increase above baseline and $g(t)$ is the normalized DA signal defined in Eq.\eqref{eq:Bt}. All signals used $G \in [50,600]$ nM in 50 nM increments, with baseline DA set to 100 nM; hereafter we report $G$ as percentage above baseline ($G_{\%}$) for ease of interpretability (see \cite{liu_detecting_2021}).

In the activated voxels we chose similar values for the magnitude and timing parameters to those in \cite{liu_detecting_2021} but lowered the magnitude of DA release in order to simulate more subtle responses:
\begin{align}
    \text{Low magnitude: }G  &\sim\mathcal{N}(150\%, (20\%)^2),\nonumber\\
    \text{High magnitude: }G &\sim\mathcal{N}(350\%, (20\%)^2),\nonumber\\
    \text{Early peak time: }t_P &<45\text{ min},\nonumber\\
    \text{Late peak time: }t_P &>45\text{ min},\nonumber\\
    \text{Time of onset: }t_D &\sim\mathcal{U}[30,40]\text{ min},\nonumber\\
    \alpha&\sim\mathcal{N}(0.7,0.1^2),
    \label{eq:prior1}
\end{align}




As preparation for the experiment, we applied the vPET-ABC with $N=10,000,000$ simulations and with a decreasing number of accepted samples $n\in\{200, 150, 100, 50, 25, 15\}$. We calculated the sensitivity and specificity based on the percentage of correctly identified activated and NULL cases, respectively, overall as well as for the different sub-groups of activated cases (i.e. “low”, “high”, “early”, “late” and their different combinations). An optimal $n$ (desired number of accepted simulations) for model selection was then determined for a further in-human study.

We then applied the vPET-ABC pipeline to a dynamic scan of a human subject smoker. Following 1 week under placebo patch as opposed to a nicotine patch and then overnight abstinence, the smoker participated in a 90-minute [$^{11}$C]raclopride scan in the HRRT PET scanner (Siemens Healthineers, Knoxville, TN) and smoked a cigarette while in the scanner starting at 35 min post injection and lasting for approximately 3 minutes. The study was approved by the Yale Human Investigation and Radiation Safety Committees. More details on the data acquisition can be found in Zakiniaeiz~\textit{et al.} \cite{zakiniaeiz_nicotine_2022}.

A 1004-voxel mask delineating the precommissural striatum and its subregions (ventral striatum, dorsal caudate, dorsal putamen), following Martinez~{et al} \cite{Martinez}, was applied prior to kinetic analysis. Within this mask, we compared the number and location of voxels identified as activated by two time-varying methods: the lp-ntPET model with effective number of parameters (ENP), as described in \cite{morris_modeling_2024,liu_model_2020} and the proposed vPET-ABC method with parameters optimized in the aforementioned simulation. The lp-ntPET ENP method produces a binary map of activated voxels while vPET-ABC generates a map of probabilities for selecting lp-ntPET over MRTM (i.e. activation vs NULL).

\subsubsection{Human [$^{18}$F]FDG TB-PET Study}
\label{real_FDG}

A $54$ year old male subject with a pancreatic neuroendocrine tumour underwent a $50$-min dynamic total-body PET scan on a Siemens Biograph Vision Quadra (Siemens Healthineers, Knoxville, TN) commencing with intravenous administration of $150$ MBq of [$^{18}$F]FDG. This scan was part of a larger study approved by the Northern Sydney Local Health District Human Research Ethics Committee. Prior written informed consent to participate in this study was obtained from the subject.

List-mode data were rebinned into 35 temporal frames (matrix size $220\times220\times645$). Standard corrections (normalization, randoms, attenuation, decay, and scatter) were applied, and images reconstructed using ordinary-Poisson ordered-subsets expectation maximization (OP-OSEM; $8$ iterations, $5$ subsets) with time-of-flight weighting and a maximum ring difference of $322$. Spatiotemporal HYPR filtering was then applied to the dynamic series \cite{christian_dynamic_2010}, prior to kinetic modeling. The arterial input function was extracted from the ascending aorta: a 3D aortic mask was generated on the native CT using MOOSE segmentation \cite{sundar_fully_2022} and eroded by two voxels to mitigate partial-volume contamination; whole-blood image-derived arterial input function (IDIF) values were defined as the frame-wise mean activity concentration within the eroded region. 

vPET-ABC was then run on the dataset using the 2TCM with model selection for the presence or absence of reversibility (i.e., $k_4=0$ or $k_4\neq0$) under wide prior distributions defined as:
\begin{align}
    K_1&\sim\mathcal{U}(0.001, 1), \quad k_2\sim\mathcal{U}(0.001, 2),\nonumber\\
    k_3&\sim\mathcal{U}(0.001, 0.5), \quad k_4\sim\mathcal{U}(0, 0.1),\nonumber\\
    V_b&\sim\mathcal{U}(0.03, 0.2), \quad m \sim Bern\left(\frac{1}{2}\right),
    \label{eq:prior2}
\end{align}
where $Bern(p)$ stands for a Bernoulli distribution of parameter $p$. 

The prior distributions were based on values reported in the literature for [$^{18}$F]FDG in a wide variety of organs and tissues \cite{yamasaki_first_2018,bullich_evaluation_2020,tan_ultralow-dose_2023}. The simulation size $N$ was selected to be sufficiently large but still fit within GPU memory and in reasonable time. The accepted number $n$ was chosen to be an arbitrarily small number to minimize bias under fixed $N$ but also not too small that Monte Carlo error is large. Thus, we selected $N=10,000,000$ and $n=18$. For more rigorous studies, we recommend that users optimize $n$ with the guidance of a pilot study, as discussed in Section.~\ref{pipeline}.

Point estimates of the kinetic parameters were calculated as the means of their respective posterior distributions conditioned on the preferred model for each voxel. A model is preferred when more than $50\%$ of all $n$ accepted simulations in a voxel is produced by the candidate model. Afterwards, a $K_i$ parametric image was created by computing $K_i=K_1k_3/(k_2+k_3)$ based on accepted samples forming $k_i$ posterior distributions, the mean estimator of which was then further compared with both a standard Patlak \cite{patlak_graphical_1983} image, and the standard SUV image as a clinical reference.


Local spatial autocorrelation preservation was assessed using local Moran's I on coronal 2D maps (SUV, Patlak $K_i$, NNLS $K_i$, and ABC $K_i$). To remove large-scale spatial trends, each map was detrended by subtracting an anisotropic Gaussian-smoothed field with FWHM 40 mm. The corresponding kernel widths were converted using the in-plane voxel spacing $(1.6456 \times 3.3\ \mathrm{mm})$. The detrended maps were then z-scored within a common mask. Local Moran's I was computed using an 8-neighbor, distance-weighted, row-standardized neighborhood. We summarized each method using the distribution of Moran's I values and the voxelwise correlations between Moran's I maps.

\section{Results and Discussion}

\subsection{2TCM Simulations}

\begin{figure*}[!t]
    \centerline{\includegraphics[width=\textwidth]{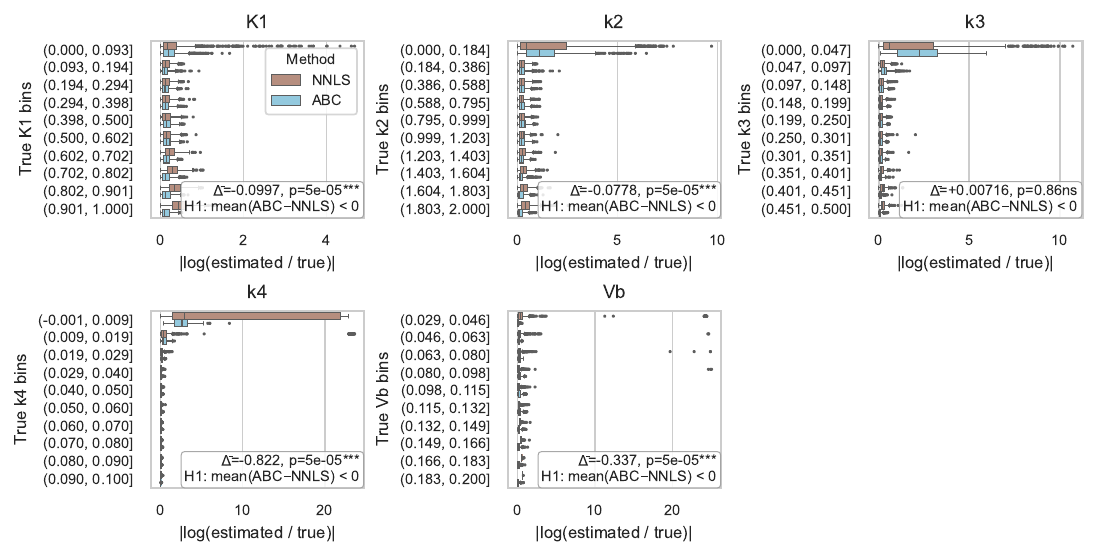}}
    \caption{Parameter point estimate accuracy comparison on the entire simulation set, comparing NNLS and vPET-ABC. The plot shows log-ratio error by true magnitude for all kinetic parameters. Paired permutation test results are shown in boxed inserts.}
    \label{fig:2tcm}
\end{figure*}

Fig.~\ref{fig:2tcm} shows distributions of log-ratio errors for all kinetic parameters, with each parameter binned by groups of true magnitude. Paired permutation tests indicate vPET-ABC performed significantly better in terms of log-ratio error of all kinetic parameters except $k_3$ which was estimated equally well by either method. For larger kinetic parameter values, vPET-ABC showed smaller bias than NNLS.

NNLS methods impose assumptions such as predefined basis grids and passbands, non-negativity constraints and time-weighting, all of which act as implicit regularization and potentially lead to biased estimates \cite{escobar_bias_1986}. This is not unique to 2TCM, as it is also evident in other methods such as lp-ntPET\cite{normandin_linear_2012} where the regressor is correlated with the regression error. In practice, NNLS methods are sensitive to observational noise, reducing sensitivity to subtle parameter differences.

\subsection{Lp-ntPET Simulations:}


\begin{table}[t]
    \begin{threeparttable}
    \caption{Runtime comparison per voxel}
    \label{tab:speed}
    \setlength\tabcolsep{0pt}
    
    \begin{tabular*}{\columnwidth}{@{\extracolsep{\fill}} ll r r r r}
    \toprule
    \multicolumn{2}{c}{} & \multicolumn{2}{c}{PC\tnote{e} Time (s)} & \multicolumn{2}{c}{HPC\tnote{f} Time (s)} \\
    \cmidrule(lr){3-4}\cmidrule(lr){5-6}
    Method & Algorithm & CPU & GPU & CPU & GPU \\
    \midrule
    NNLS+Basis Library & FNNLS\tnote{a} & 0.181 & \textemdash & $<$0.0001 & 0.075 \\
    SMC & Independent MH\tnote{b} & 58 & \textemdash & \textemdash & \textemdash \\
    HMC & NUTS\tnote{c} & 137 & 16050 & 462 & 4928 \\
    ABC & vPET-ABC\tnote{d} & \textemdash & 0.239 & \textemdash & 0.0065 \\
    \bottomrule
    \end{tabular*}
    
    \smallskip
    \scriptsize
    \begin{tablenotes}
    \RaggedRight
    \item[a] Fast Non-negative Least Squares \cite{bro_fast_1997} \texttt{scipy.optimize.nnls} for CPUs and \texttt{jaxfit.CurveFit} for GPUs. The basis library size was 1185; not vectorized.
    \item[b] Independent Metropolis-Hastings, \texttt{pymc.smc.sample\_smc}. $10{,}000$ draws.
    \item[c] No-U-Turn Sampler, \texttt{pymc.sampling.jax.sample\_numpyro\_nuts}. $5{,}000$ burn-in and $10{,}000$ draws.
    \item[d] \texttt{vpetabc.ABCRejection}.
    \item[e] Commodity PC, CPU: 12th Gen Intel(R) Core(TM) i7-12700H (2.30 GHz), GPU: NVIDIA GeForce RTX 3060 Laptop GPU. JAX does not support NVIDIA CUDA on Windows x64, all code run under Windows WSL2.
    \item[f] NCI Gadi cluster, CPU: 24-core Intel Xeon Scalable 'Cascade Lake', GPU: NVIDIA V100.
    \end{tablenotes}
    \end{threeparttable}
\end{table}

\begin{table}[t]
    \begin{threeparttable}
    \caption{Model-selection performance by noise level}
    \label{tab:model_accuracy}
    \setlength\tabcolsep{0pt} 
    
    \begin{tabular*}{\columnwidth}{@{\extracolsep{\fill}} l l r r}
    \toprule
    \multicolumn{2}{c}{} & \multicolumn{2}{c}{Performance} \\
    \cmidrule(lr){3-4}
    Method & Noise & Sensitivity\tnote{a} & Specificity \\
    \midrule
    NNLS+Basis Library\tnote{b}     & Low  & 0.843 & 0.994 \\
    ABC & Low  & 0.925 & 0.955 \\
    \addlinespace
    NNLS+Basis Library     & High & 0.102 & 0.996 \\
    ABC & High & 0.880 & 0.575 \\
    \bottomrule
    \end{tabular*}
    
    \smallskip
    \scriptsize
    \begin{tablenotes}
    \RaggedRight
    \item[a] Positive = activation ($\gamma\neq 0$ in lp-ntPET).
    \item[b] NNLS+Basis Library model selection via Bayesian information criteria (BIC).
    \end{tablenotes}
    \end{threeparttable}
\end{table}

\begin{figure*}[!t]
    \centerline{\includegraphics[width=\textwidth]{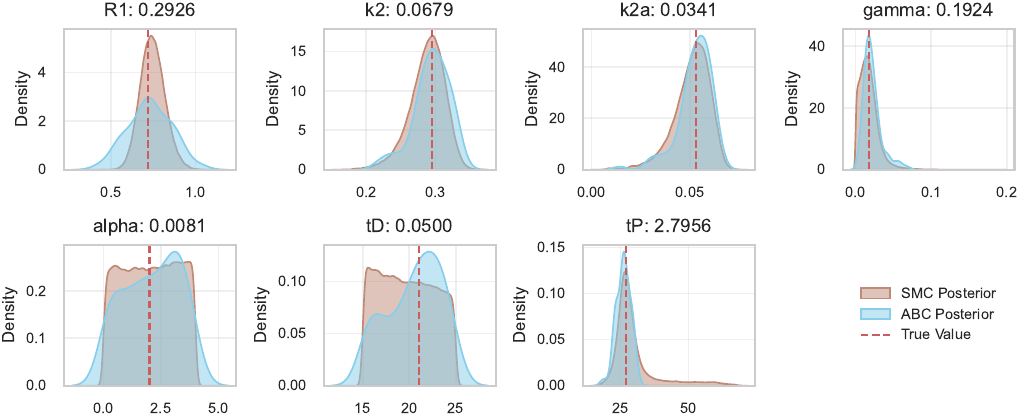}}
    \caption{Posterior distribution comparison between SMC and vPET-ABC on a TAC with activation. The values shown in each subplot title are the KL values: $\mathrm{KL}[\text{SMC}\mid\text{ABC}]$. The red dashed vertical lines indicate true parameter values.} 
    \label{fig:posteriors}
\end{figure*}

Table~\ref{tab:speed} compares runtime for four methods: NNLS+Basis Library, Monte Carlo baselines (SMC and HMC/NUTS) and vPET-ABC. vPET-ABC on GPUs achieves substantially lower runtime than SMC/HMC and is comparable to NNLS on CPUs, indicating effective use of accelerator hardware. In contrast, the GPU implementation of NNLS was slower than its CPU counterpart in our experiments, suggesting limited parallelization efficiency in this setting and less effective use of GPU hardware than vPET-ABC.

Fig.~\ref{fig:posteriors} depicts the accuracy of posterior distribution estimation by vPET-ABC relative to SMC which is treated as the ground truth. An arbitary TAC was chosen for demonstration as all TACs are different noise realizations of the same physiologically feasible TAC. Convergence of the SMC posterior approximation was assessed using four independent SMC runs (PyMC \texttt{sample\_smc}, \texttt{chains}=4). We evaluated between-run agreement using Gelman--Rubin $\hat{R}$ and effective sample size (ESS) computed in ArviZ. All parameters had $\hat{R}\le 1.01$ (mode $1.00$), with adequate ESS, indicating stable posterior estimates across independent SMC runs. Most vPET-ABC posteriors resemble those of SMC with small KL divergences, with the minor exception of $R_1$ where SMC produced a posterior distribution with a narrower spread, $t_D$ where SMC revealed a roughly monotonically decreasing distribution, and $t_P$ where SMC identify an asymmetric distribution with a heavy upper tail.

For model selection sensitivity, ABC and NNLS with a basis library perform similarly at the lower noise level (Table~\ref{tab:model_accuracy}). At the higher noise level, ABC maintains higher sensitivity with moderate specificity. While NNLS model selection is often implemented via F-tests or information criteria (IC) \cite{golla_model_2017}, these approaches have well-known pitfalls in kinetic modeling. AIC penalizes model complexity using only the number of free parameters, whereas the effective complexity also depends on parameter identifiability (i.e., the local information of the likelihood) \cite{cavanaugh_akaike_2019}. AIC is easily biased by boundary constraints \cite{cavanaugh_akaike_2019}, and often requires manual calibration to improve sensitivity \cite{liu_model_2020}. Moreover, many IC are plug-in scores evaluated at point estimates, discarding posterior uncertainty \cite{gelman_understanding_2014}, and even defining the effective sample size for penalization is non-trivial given temporally correlated TACs and implementation details such as up-sampled convolution or basis-function parameterization \cite{liu_model_2020}. Finally, the “linear” lp-ntPET formulation (Eq.~\eqref{eq:lp-ntPET}) includes a regressor $\int_0^t C_T(u),du$ that can be correlated with noise, violating Gauss–Markov assumptions (notably strict exogeneity), so least-squares estimates may not be BLUE (best linear unbiased estimates) and may yield biased parameter estimates and model selection.

Note that, as our pilot run was intended to optimize ABC hyperparameters for better kinetic parameter accuracy, suboptimal model selection accuracy was expected. For a study focusing on model selection (such as Section.~\ref{real_smoking}), a different pilot study designed to optimize model sensitivity or specificity can improve the corresponding result further.

\subsection{Human Cigarette Smoking Study}

In Table~\ref{tab:phantom_detection}, we report the percentage of correctly detected activated voxels and null voxels. For the null case, the percentage decreases as the number of accepted simulations ($n$) is reduced, ranging from $99.5\%$ to $89.2\%$. In contrast, for the activated voxels, the percentage increases over the same range, from $88.8\%$ to $99.2\%$. Notably, the two percentages converge at 50 accepted simulations. Also, for the sub-categories of activated voxels, the lowest sensitivity is observed in the “low” magnitude condition, where detection drops to $79.5\%$ with $200$ accepted simulations $n$. Within this category, the “early–low” case performs worst, reaching only $71.7\%$ accuracy at the same tolerance level. By comparison, the “high” magnitude case consistently yields near-perfect performance, remaining at $100\%$ across most conditions, with only a slight decrease to $99.8\%$ in the “high–early” case at $150$ and $200$ accepted simulations $n$.

\begin{table*}[t]
\centering
\begin{threeparttable}
\caption{Detection performance for 4D Phantom vPET-ABC with $N=10{,}000{,}000$}
\label{tab:phantom_detection}

\setlength\tabcolsep{3pt}
\footnotesize

\begin{tabular*}{\textwidth}{@{\extracolsep{\fill}} l r r r r r r r r r r}
\toprule
 Choice of $n$ &
 Activated &
 NULL &
 Activated &
 Activated &
 Activated &
 Activated &
 Activated early &
 Activated early &
 Activated late &
 Activated late \\
 & voxels &
 voxels &
 low &
 high &
 early &
 late &
 and low &
 and high &
 and low &
 and high \\
\midrule
Detected at $n=200$
 & 88.80\% & 99.50\% & 79.50\% & 99.80\% & 85.70\% & 96.80\% & 71.70\% & 99.80\% & 95.00\% & 100.00\% \\
Detected at $n=150$
 & 91.20\% & 98.50\% & 84.00\% & 99.80\% & 88.50\% & 98.00\% & 77.00\% & 99.80\% & 97.00\% & 100.00\% \\
Detected at $n=100$
 & 93.80\% & 98.00\% & 88.50\% & 100.00\% & 91.70\% & 99.20\% & 83.50\% & 100.00\% & 98.70\% & 100.00\% \\
Detected at $n=50$
 & 96.80\% & 96.80\% & 94.00\% & 100.00\% & 95.80\% & 99.20\% & 91.70\% & 100.00\% & 98.70\% & 100.00\% \\
Detected at $n=25$
 & 98.50\% & 93.00\% & 97.00\% & 100.00\% & 98.00\% & 99.50\% & 96.00\% & 100.00\% & 99.00\% & 100.00\% \\
Detected at $n=15$
 & 99.20\% & 89.30\% & 98.50\% & 100.00\% & 98.80\% & 100.00\% & 97.70\% & 100.00\% & 100.00\% & 100.00\% \\
\bottomrule
\end{tabular*}

\smallskip
\scriptsize
\begin{tablenotes}
\RaggedRight
\item Detection rates are reported as the percentage of correctly identified voxels for each activation class.
\end{tablenotes}
\end{threeparttable}
\end{table*}

\begin{figure}[!t]
    \centering
    \includegraphics[width=\columnwidth]{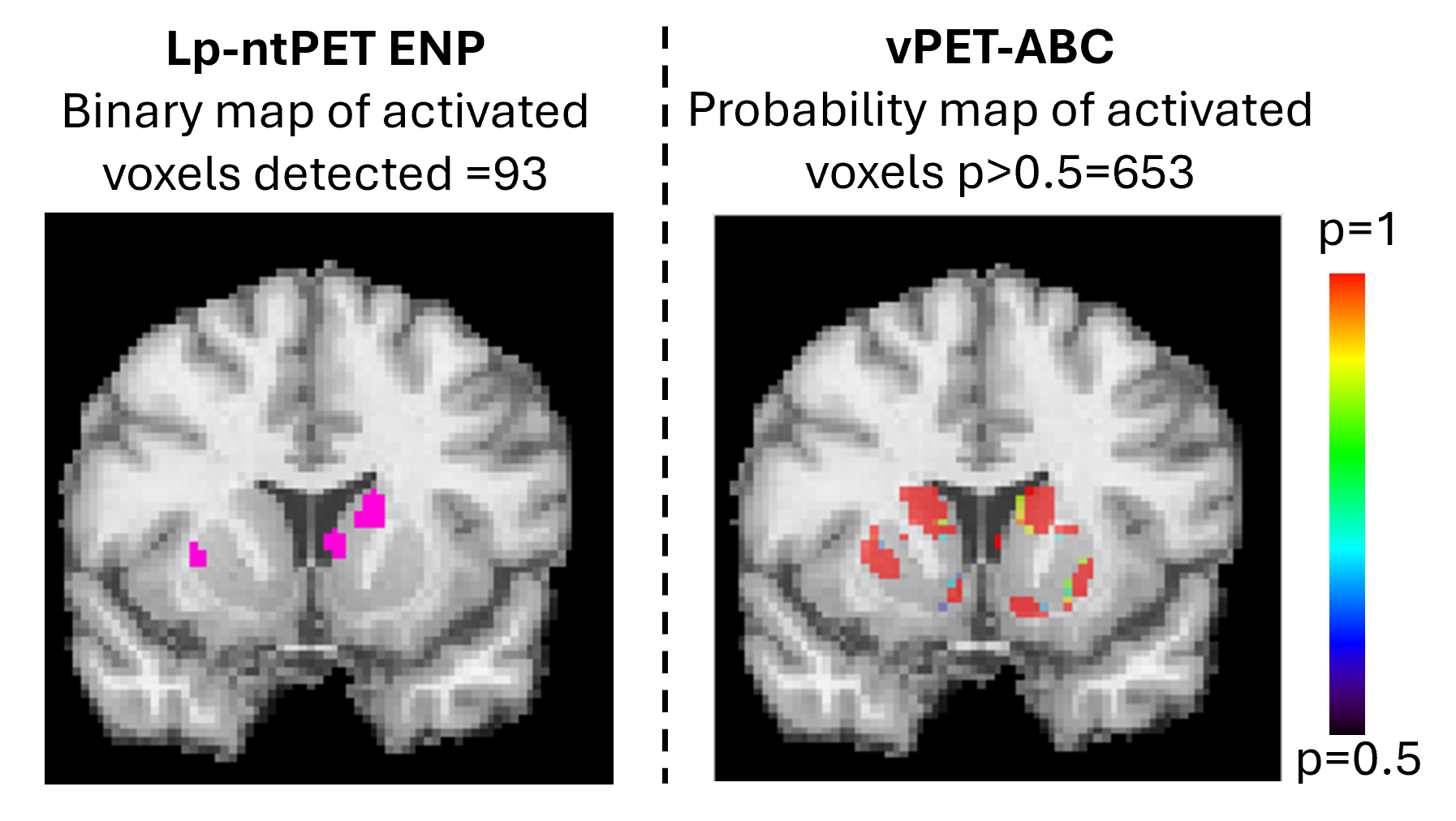}
    \caption{Model selection maps from a human [\textsuperscript{11}C]raclopride PET dataset co-registered to the MNI template. Left: binary lp-ntPET ENP map, where pink voxels indicate significant dopamine activation. Right: vPET-ABC activation probability map, with higher probabilities ($p \in [0.5,1]$) indicating stronger evidence for activation.}
    \label{fig:6}
\end{figure}


For the human smoking dataset, we conducted a proof-of-concept analysis comparing vPET-ABC to lp-ntPET with ENP. In Fig.~\ref{fig:6}, detected activations are shown as a binary map for the lp-ntPET ENP method (in pink) and as a map of model probability (lp-ntPET versus MRTM) for the vPET-ABC method, with $\text{posterior model probability} > 0.5$ indicating that an activation is more likely than not. In this analysis, we adopted a more conservative strategy than in the simulation study. While $n = 50$ (out of $N = 10{,}000{,}000$ simulations) achieved the best overall performance, we instead chose $n = 100$ to prioritize control of false positives while preserving an acceptable false-negative rate (Table~\ref{tab:phantom_detection}). The results show a wider extent of significant dopamine activation using vPET-ABC than lp-ntPET ENP, suggesting improved sensitivity of vPET-ABC. Furthermore, a substantial fraction of voxels in the vPET-ABC probability map have a probability of $1$, indicating a maximal degree of certainty of activation at this tolerance level.

\subsection{Human [$^{18}$F]FDG TB-PET Study}

 \begin{figure*}[!t]
    \centerline{\includegraphics[width=\textwidth]{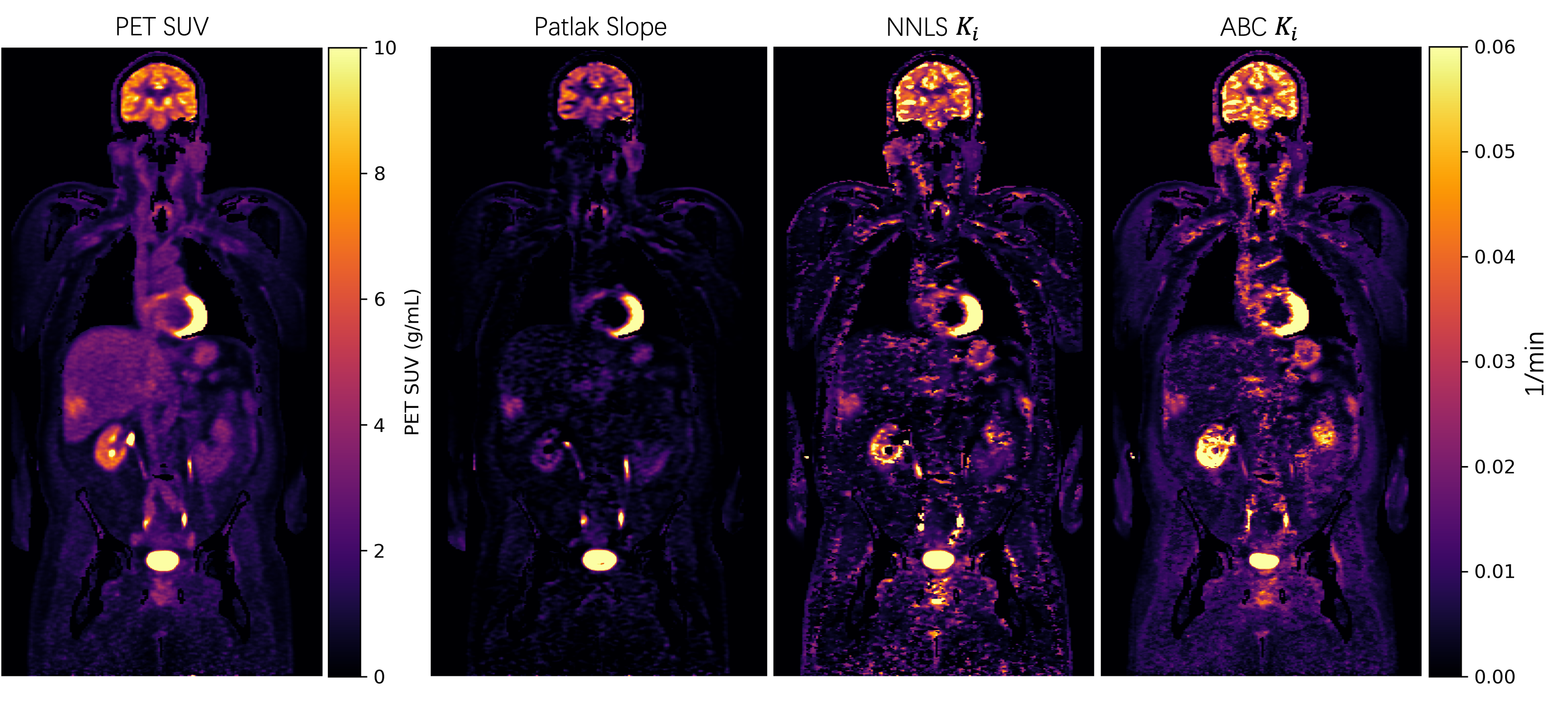}}
    \caption{SUV image (left) and $K_i$ images obtained using (from left to right): Patlak plot, non linear least squares with non-negativity constraints (NNLS) and vPET-ABC. No post-processing smoothing filter was applied.}
    \label{fig:TB-PET_result}
\end{figure*}

Our vPET-ABC pipeline successfully computed kinetic parameter posterior distributions for all $4.4$ million voxels in the TB-PET dataset, within $2$ minutes for a single coronal slice and $10$ hours for the whole volume using the NCI Gadi cluster \cite{nci_gadi_2019} with a single NVIDIA V100. This is comparable to our in-house NNLS basis function implementation. The computation time was much faster than the anticipated $36667$ hours with b-ntPET \cite{irace_bayesian_2020}, $733333$ hours with iDDPM deep neural network \cite{djebra_bayesian_2025}, or $733333$ hours with SMC. vPET-ABC has a significant speed advantage over other methods, making it an appealing choice for Bayesian kinetic modeling in large volumes of data. While the $10$ hours computation time is still less than ideal for practical clinical use, it is worth noting a trade-off between estimation quality and computational cost can be easily made by adjusting the simulation size, achieving orders of magnitude speed-up.

As illustrated in Fig.~\ref{fig:TB-PET_result}, there is good agreement between the vPET-ABC $K_i$ parametric image and the NNLS $K_i$ image. In particular, the hyper-metabolic tumor in the right lower lobe of the liver is observed in both images. The Patlak image aligns well with the other $K_i$ images in the brain, myocardium and liver tumor. However, Patlak underestimates $K_i$ in other tissues where signal-to-noise is low (e.g. due to low tracer uptake) or the TACs do not comply with the assumption of irreversibile kinetics \cite{karakatsanis_generalized_2015}.

The vPET-ABC $K_i$ parametric image also appears to have less voxel to voxel variability than NNLS, likely because the Bayesian formulation regularizes voxelwise inference (via the prior and posterior averaging), reducing noise amplification and yielding more spatially coherent parameter maps that reflect the locally spatial-correlated TACs. For a quantitative measure, after detrending each parametric map by subtracting a heavily smoothed (Gaussian FWHM 40 mm) low-frequency background to remove global spatial gradients, and z-scoring within a common mask, local Moran’s $I$ differed across maps. The upper-tail clustering (q75; the 75th percentile of voxelwise local Moran’s $I$) was highest for vPET-ABC (0.0218), followed by SUV (0.0115), NNLS (0.00848), and Patlak (0.00506). Here larger positive $I$ indicates stronger local clustering of similar voxel values (greater spatial coherence), values near zero indicate weak local association, and negative values indicate local spatial outliers, while medians were near zero for all methods. Spatial agreement of local autocorrelation structure (Pearson correlation between voxelwise local Moran’s $I$ maps) was strongest between vPET-ABC $K_i$ and SUV (r = 0.8546), moderate for NNLS $K_i$ and SUV (r = 0.6402), and lowest for Patlak $K_i$ and SUV (r = 0.4805). We used SUV as a semi-quantitative reference because it is the most widely used clinical measure and typically exhibits the expected spatial coherence of reconstructed uptake images under standard preprocessing. Thus, agreement with local autocorrelation patterns in the SUV image provides a pragmatic reference for assessing whether a $K_i$ estimation method yields spatially plausible parametric maps and preserves SUV-like spatial texture.

\section{Conclusions and Limitations}

\label{sec:conclusions}
We have introduced a fully vectorized voxelwise ABC framework for next generation PET kinetic modeling that scales to total-body data and returns voxelwise posteriors, model probabilities, and point estimates without requiring an explicit or differentiable likelihood function. In simulations, posterior means were significantly more accurate than NNLS and posterior distributions closely matched SMC references. For model selection between lp-ntPET and MRTM, vPET-ABC maintained higher sensitivity under high noise at some cost to specificity, whereas NNLS with BIC showed the opposite trade-off with an extremely low sensitivity. In a proof-of-concept smoking dataset, vPET-ABC produced probabilistic maps in broad agreement with lp-ntPET with ENP but with more extensive activation detected, at a conservative tolerance. In a proof-of-concept total-body [$^{18}$F]FDG dataset, vPET-ABC generated $K_i$ maps and model probability maps (not shown) in practical runtimes on a single data-center GPU, in good agreement with NNLS. Consistent with this, detrended local Moran’s I indicated that vPET-ABC preserved SUV-like local spatial dependence more closely than NNLS and Patlak.

For practical use pipelines we recommend full-TAC summaries and an $L_1$ distance, with the acceptance proportion selected by a short, tracer-matched pilot. The main engineering constraint is GPU memory, so batching simulations is recommended (automated in our released package), and preprocessing should be consistent between pilot simulations and analysis. A realistic 4D striatal phantom confirmed that tuning the acceptance ratio offers a practical trade-off between sensitivity and specificity across activation magnitude and timing. 

Limitations include the approximate nature of ABC posteriors. Future work should account for spatial correlations in the simulation, and benchmark against the latest deep learning kinetic modeling methods \cite{zhao_generative_2026}. Overall, vPET-ABC provides fast, accurate, training-free likelihood-free Bayesian inference for next-generation PET kinetics with uncertainty quantification that is practical at the whole-body scale.

\section*{Code Availbility}

The \texttt{vpet-abc} package is available in https://github.com/zephyralistair/vPET-ABC-fast-PET-kinetic-modeling-on-large-data/tree/main and can be installed via \texttt{pip}.











\section*{Acknowledgment}
The authors would like to thank Paul Roach, Dale Bailey, Sally Ayesa and Liz Bailey for their advice on protocol design and the nuclear medicine staff at Royal North Shore Hospital for performing the TB-PET studies. We also acknowledge the facilities and the scientific and technical assistance of Sydney Imaging, a core research facility at The University of Sydney, and the National Imaging Facility, a National Collaborative Research Infrastructure Strategy (NCRIS) Capability. This research was also undertaken with the use of the National Computational Infrastructure (NCI Australia) \cite{nci_gadi_2019}. NCI Australia is enabled by the National Collaborative Research Infrastructure Strategy (NCRIS).

\bibliographystyle{IEEEtran}
\bibliography{tmi}

\end{document}